\documentclass[doublecol,linenumbers]{epl2} 

\usepackage{txfonts}
\usepackage{bm}
\def\rnum#1{\expandafter{\romannumeral #1}} 
\def\Rnum#1{\uppercase\expandafter{\romannumeral #1}}

\title{Magnetic anisotropy of critical current in nanowire Josephson junction with
spin-orbit interaction}
\shorttitle{Magnetic anisotropy of critical current in NW JJ with SOI}

\author{T.\ Yokoyama\inst{1,2} \and Yu.\ V.\ Nazarov\inst{1}}
\shortauthor{T.\ Yokoyama \etal}

\institute{                    
  \inst{1} Kavli Institute of Nanoscience, Delft University of Technology,
Lorentzweg 1, 2628 CJ, Delft, The Netherlands \\
  \inst{2} Center for Emergent Matter Science, RIKEN institute,
2-1 Hirosawa, Wako, Saitama 351-0198, Japan
}

\abstract{
We develop and study theoretically a minimal model of semiconductor
nanowire Josephson junction that incorporates Zeeman and spin-orbit
effects. The DC Josephson current is evaluated from the phase-dependent
energies of Andreev levels. Upon changing the magnetic field applied,
the critical current oscillates manifesting cusps that signal the $0$-$\pi$
transition. Without spin-orbit interaction, the oscillations and positions of
cusps are regular and do not depend on the direction of magnetic field.
In the presence of spin-orbit interaction, the magnetic field dependence of
the current becomes anisotropic and irregular. We investigate this
dependence in detail and show that it may be used to characterize
the strength and direction of spin-orbit interaction in experiments with
nanowires.
}

\begin{document}
\maketitle

\section{Introduction}

Semiconductor nanowire is an attractive nanostructure to investigate
spin physics arising from the spin-orbit (SO) interaction.
A strong SO interaction and the manipulation of electron spin in
InAs and InSb nanowires have been reported~\cite{Nadj-Perge1,Nadj-Perge2}.
Such nanowires are candidates to realize the topological physics.
It has been suggested that the superconductor-nanowire junction
forms the Majorana fermion at edge of superconducting region~\cite{Kitaev}.
The zero-bias anomaly of conductance, which is attributed to
the Majorana bound stats, has been measured in the transport
experiments for InSb nanowire recently~\cite{Mourik,Das,Deng}.
The nanowire Josephson junctions have been also examined beyond
topologically non-trivial regime~\cite{Doh,Dam,Nilsson,Rokhinson,Gharavi,Li,private}.

The Josephson effect is one of the most fundamental phenomena in
superconductivity. The spin degree of freedom enriches physics of
the Josephson effect, e.g., causing the $0$-$\pi$ transition in
ferromagnetic Josephson junction~\cite{Oboznov}.
In recent studies, the parity conservation of quasiparticles was
shown to cause the $4\pi$-periodicity of current-phase relation~\cite{Fu}
and increase of critical current~\cite{Beenakker1}.
The effect of SO interaction on the Josephson effect also
has been studied. The SO interaction in combination with
magnetic field shifts the current-phase relation, which results in
the anomalous Josephson current that persists even at zero
phase difference~\cite{Buzdin,YEN1,YEN2}. In previous studies,
we have attributed the anomalous effect to the spin-dependent
channel mixing in the nanowire~\cite{YEN1,YEN2}.

In this letter, we study the magnetic field dependence of critical current
in the presence of SO interaction. We develop a minimal model that
encompasses the effect. The critical current oscillation accompanying
the $0$-$\pi$ transition has been demonstrated in ferromagnetic
Josephson junctions~\cite{Oboznov}. Recent experiment has reported
a similar oscillation in the magnetic field in InSb nanowire~\cite{private}.
The oscillation should be affected by the effective SO field.
The SO interaction has been discussed as the origin of anisotropies of
level anticrossing, g-factor etc~\cite{Nadj-Perge1,Nadj-Perge2,Takahashi}.
We investigate the anisotropy of critical current magnetic field dependence.
The distance between the cusps of critical current is modulated by
the angle between magnetic and effective SO field.
If the SO interaction is strong, the two cusps are closer to each other.
From the measurement of the anisotropy of critical current~\cite{private},
the direction of the effective SO field can be determined.
We also discuss the parity effect on the critical current oscillation.

\section{Model}

\begin{figure}
\includegraphics[width=88mm]{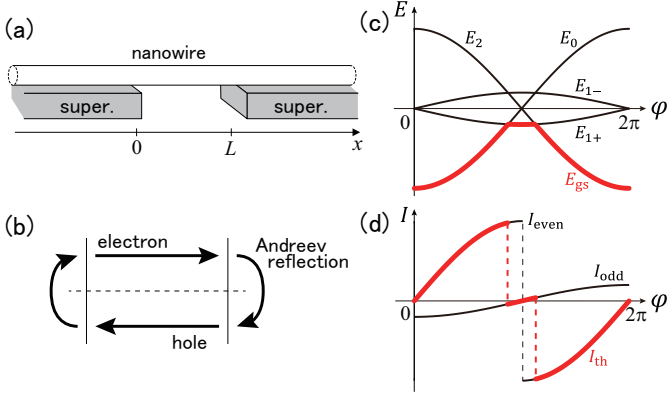}
\caption{
Minimal model of a semiconductor nanowire Josephson junction.
(a) Schematic view of the model. The nanowire is infinitely long
along the $x$ axis. The electron and hole transport in the ballistic
nanowire. The pair potential is induced in the nanowire at $x<0$ and
$x>L$ by the proximity effect.
(b) Schematic view of Andreev bound state formed by
right-going electron and left-going hole.
(c) Energies of junction without quasiparticles and with one or two
quasiparticles as functions of phase difference $\varphi$.
The two middle curves correspond to one quasiparticle
states, the levels of which are split by a weak magnetic field.
Red thick curve corresponds to the ground state.
(d) Josephson currents in even and odd parity states at zero
temperature (black lines) and in the ground state at finite
temperature (red thick line).
}
\label{fig:Model}
\end{figure}

Let us formulate a minimal model.
We consider an one-dimensional semiconductor nanowire
connected to two superconductors, as shown in Fig.\ \ref{fig:Model}(a),
that supports a single transport mode.
The nanowire has no impurities and is infinitely long along the $x$ axis.
Therefore, the electron and hole propagations are completely ballistic.
The spin singlet superconducting pair potential $\Delta (x)$ is
induced in the nanowire by the proximity effect.

The Bogoliubov-de Gennes (BdG) equation reads as~\cite{NazarovBlanter}
\begin{equation}
\left( \begin{array}{cc}
H - E_{\rm F} & \Delta (x) \\
\Delta^* (x)   & -(\mathcal{T} H \mathcal{T}^{-1} - E_{\rm F})
\end{array} \right)
\left( \begin{array}{c}
\bm{\psi}_{\rm e} \\
\bm{\phi}_{\rm h}
\end{array} \right)
= E \left( \begin{array}{c}
\bm{\psi}_{\rm e} \\
\bm{\phi}_{\rm h}
\end{array} \right).
\label{eq:BdG}
\end{equation}
Here $\bm{\psi}_{\rm e} = (\psi_{{\rm e} +}, \psi_{{\rm e} -} )^{\rm T}$
and $\bm{\phi}_{\rm h} = (-\phi_{{\rm h} -}, \phi_{{\rm h} +} )^{\rm T}$
are the spinors for electron and hole, respectively. We assume
$\Delta (x) =\Delta_0 \{ e^{i\varphi_{\rm L}} \vartheta (-x)
+ e^{i\varphi_{\rm R}} \vartheta (x-L) \}$ with the step function
$\vartheta (t) =1$ for $t \geq 0$ and $0$ for $t<0$.
The phase difference between two superconductors is defined as
$\varphi \equiv \varphi_{\rm L} - \varphi_{\rm R}$.
The energy $E$ is measured from the Fermi level $E_{\rm F}$.
$H$ in the diagonal element is the free-electron Hamiltonian.
In our model, $H = H_0 + H_{\rm SO} + H_{\rm Z}$ with $H_0= {p_x}^2 /(2m^*)$,
the SO interaction $H_{\rm SO} = \bm{\alpha} \cdot \hat{\bm{\sigma}} p_x /\hbar$,
and the Zeeman effect due to an external magnetic field
$H_{\rm Z} = g \mu_{\rm B} \bm{B} \cdot \hat{\bm{\sigma}}/2$
using effective mass $m^*$, $g$-factor $g$ ($\simeq -50$ for InSb),
Bohr magneton $\mu_{\rm B}$, and Pauli matrices $\hat{\bm{\sigma}}$.
The time-reversal operator $\mathcal{T} = -i \hat{\sigma}_y K$
satisfies $\mathcal{T} H_{\rm SO} \mathcal{T}^{-1} = H_{\rm SO}$
and $\mathcal{T} H_{\rm Z} \mathcal{T}^{-1} = - H_{\rm Z}$.
$K$ is the operator to form a complex conjugate; $K f = f^*$.
The Hamiltonian is rewritten as
\begin{equation}
H = \frac{{p_x}^2}{2m^*} - \frac{\alpha}{\hbar} p_x \hat{\sigma}_\theta
- \frac{1}{2} E_{\rm Z} \hat{\sigma}_z
\end{equation}
by choosing proper axis in spin space. Here $\hat{\sigma}_\theta
= \hat{\sigma}_z \cos \theta + \hat{\sigma}_x \sin \theta$ with
$\theta$ being the angle between the external field and
the effective SO field. $E_{\rm Z} \equiv |g \mu_B B|$.

The magnetic field is assumed to be screened in the superconducting
region and the Zeeman energy $E_{\rm Z}$ is non-zero only at $0<x<L$.
For a large $g$-factor in InSb, a large Zeeman energy is obtained for
weak magnetic field, which does not break the superconductivity.
We assume a short junction with $L \ll \xi \equiv \hbar v_{\rm F}/(\pi \Delta_0)$.
No potential barrier is assumed at the boundaries between the normal
and superconducting regions, so that the electron propagation in
the nanowire is completely ballistic. $E_{\rm Z}$ and $\Delta_0$ are
much smaller than $E_{\rm F}$.

The BdG equation in Eq.\ (\ref{eq:BdG}) gives a pair of Andreev levels.
When the BdG equation has an eigenenergy $E_n (\varphi)$ with
eigenvector $(\bm{\psi}_{{\rm e},n}, \bm{\phi}_{{\rm h},n})^{\rm T}$,
$-E_n (\varphi)$ is also an eigenenergy of the equation with
$(-\mathcal{T} \bm{\phi}_{{\rm h},n}, \mathcal{T} \bm{\psi}_{{\rm e},n})^{\rm T}$.

\section{CALCULATION AND RESULTS}

The BdG equation in Eq.\ (\ref{eq:BdG}) can be written in
terms of the scattering matrix~\cite{Beenakker2}.
We focus on a single conduction channel in the nanowire.

Let us consider the wavefunction of the form
$(\bm{\psi}_{\rm e}, \bm{\phi}_{\rm h})^{\rm T} = e^{\pm ik_{\rm F} x}
(\bm{\psi}_{\rm e}^{(\pm)}, \bm{\phi}_{\rm h}^{(\pm)})^{\rm T}$.
The envelope function with positive (negative) sign corresponds to
the quasiparticle for right-going (left-going) electron and left-going
(right-going) hole. The BdG equation for the envelope function is
given by
\begin{equation}
\left( \hspace{-1mm} \begin{array}{cc}
\mp i \hbar v_{\rm F} \partial_x - \bm{h}_\pm \cdot \hat{\bm{\sigma}}
& \Delta (x) \\
\Delta^* (x) &
\pm i \hbar v_{\rm F} \partial_x - \bm{h}_\mp \cdot \hat{\bm{\sigma}} 
\hspace{-1mm} \end{array} \right)
\left( \hspace{-1mm} \begin{array}{c}
\bm{\psi}_{\rm e}^{(\pm)} \\
\bm{\phi}_{\rm h}^{(\pm)}
\end{array} \hspace{-1mm} \right)
= E
\left( \hspace{-1mm} \begin{array}{c}
\bm{\psi}_{\rm e}^{(\pm)} \\
\bm{\phi}_{\rm h}^{(\pm)}
\end{array} \hspace{-1mm} \right)
\label{eq:modBdG}
\end{equation}
with
\begin{equation}
\bm{h}_\pm = \frac{1}{2} E_{\rm Z} \bm{e}_z
\pm \alpha k_{\rm F} \bm{e}_\theta,
\label{eq:totalB}
\end{equation}
which means a total magnetic field for electron and hole.
Here $\partial_x^2 \psi_{\rm e}^{(\pm)}$ and
$\alpha (\partial_x \psi_{\rm e}^{(\pm)})$ terms and
those for hole are neglected when $E_{\rm F} \gg \Delta_0$.

The quantum transport of electron (hole) in the normal region
($\Delta =0$) with SO interaction and Zeeman effect is
described by the scattering matrix $S_{\rm e}$ ($S_{\rm h}$).
The wavefunctions of electron and hole are
$\bm{\psi}_{\rm e}^{(\pm)} (x) \propto \exp
(\mp i \frac{E + \bm{h}_\mp \cdot \hat{\bm{\sigma}} }{\hbar v_{\rm F}} x)$
and
$\bm{\phi}_{\rm h}^{(\pm)} (x) \propto \exp
(\pm i \frac{E + \bm{h}_\pm \cdot \hat{\bm{\sigma}} }{\hbar v_{\rm F}} x)$,
respectively.
The scattering matrices are related to each other by
$\hat{S}_{\rm h}(E) = \hat{g} \hat{S}_{\rm e}^* (-E) \hat{g}^\dagger$
with $\hat{g} \equiv -i \hat{\sigma}_y$.
On the assumption that they are independent of energy $E$
for $|E|<\Delta_0$, and thus
$\hat{S}_{\rm h} = \hat{g} \hat{S}_{\rm e}^* \hat{g}^\dagger$.
The transmission coefficient for the ballistic nanowire is unity.
We denote $\hat{S}_{\rm e} = \hat{S}$:
\begin{equation}
\hat{S} = \left( \begin{array}{cc}
 & \hat{t}_{\rm LR} \\
\hat{t}_{\rm RL} &
\end{array} \right),
\label{eq:smatrix}
\end{equation}
where
\begin{equation}
\hat{t}_{\rm RL} = \exp
\left( i \frac{L}{\hbar v_{\rm F}} \bm{h}_+ \cdot \hat{\bm{\sigma}} \right)
\hspace{2mm}, \hspace{2mm}
\hat{t}_{\rm LR} = \exp
\left( -i \frac{L}{\hbar v_{\rm F}} \bm{h}_- \cdot \hat{\bm{\sigma}} \right)
\end{equation}
mean dynamical phases for spins by SO interaction and
Zeeman effect (see supplementary note).

The Andreev reflection at $x=0$ and $L$ is also described in
terms of scattering matrix $\hat{r}_{\rm he}$ for
the conversion from electron to hole and $\hat{r}_{\rm eh}$ for
that from hole to electron~\cite{Beenakker2}:
\begin{equation}
\hat{r}_{\rm he} = e^{-i \alpha_{\rm A}}
\left( \begin{array}{cc}
 e^{-i \varphi_{\rm L}} & \\
 & e^{-i \varphi_{\rm R}}
\end{array} \right)
\hspace{2mm}, \hspace{2mm}
\hat{r}_{\rm eh} = e^{-i \alpha_{\rm A}}
\left( \begin{array}{cc}
 e^{i \varphi_{\rm L}} & \\
 & e^{i \varphi_{\rm R}}
\end{array} \right)
\label{eq:AR}
\end{equation}
with $\alpha_{\rm A} \equiv \arccos(E /\Delta_0)$. It is important
that $\hat{r}_{\rm he (eh)}$ does not depend on SO interaction.

The product of scattering matrices gives an equation,
$\det \left( \hat{1} - \hat{r}_{\rm eh} (\hat{g} \hat{S}^* \hat{g}^\dagger )
\hat{r}_{\rm he} \hat{S} \right) = 0$,
which is equivalent with the BdG equation in Eq.\ (\ref{eq:BdG}),
and gives the energies of Andreev levels.
By substituting Eqs.\ (\ref{eq:smatrix}) and (\ref{eq:AR}), we obtain
\begin{eqnarray}
\det \left( \hat{1} - e^{ -i (2\alpha_{\rm A} - \varphi)}
\hat{t}_{\rm LR} \hat{t}_{\rm RL} \right) &=& 0,
\label{eq:detCW} \\
\det \left( \hat{1} - e^{ -i (2\alpha_{\rm A} + \varphi)}
\hat{t}_{\rm RL} \hat{t}_{\rm LR} \right) &=& 0.
\label{eq:detCCW}
\end{eqnarray}
Eq.\ (\ref{eq:detCW}) corresponds to the Andreev bound states with
clockwise path (right-going electron and left-going hole) schematically
shown in Fig.\ \ref{fig:Model}(b), whereas Eq.\ (\ref{eq:detCCW}) is
a counterclockwise path. The energies of Andreev level are given by
\begin{eqnarray}
E_{\circlearrowright ,\pm} (\varphi ) &=& \Delta_0
\cos \left( \frac{\varphi \pm \tilde{\theta}}{2} \right), \\
E_{\circlearrowleft ,\pm} (\varphi ) &=& \Delta_0
\cos \left( \frac{-\varphi \pm \tilde{\theta} + 2\pi}{2} \right)
= - E_{\circlearrowright ,\mp} (\varphi )
\end{eqnarray}
for clockwise and counterclockwise paths, respectively.
The phase $\tilde{\theta} \in [0,\pi]$ is defined as
\begin{equation}
\cos \tilde{\theta} \equiv \cos \theta_+ \cos \theta_-
- (\bm{n}_+ \cdot \bm{n}_-) \sin \theta_+ \sin \theta_-
\label{eq:phase}
\end{equation}
with $\theta_\pm = |\bm{h}_\pm| L/(\hbar v_{\rm F})$~\cite{com}.
$\bm{n}_\pm = \bm{h}_\pm / |\bm{h}_\pm|$ are unit vectors.
We introduce parameters $\theta_B \equiv E_{\rm Z} L/(\hbar v_{\rm F})$
and $\theta_{\rm SO} \equiv 2\alpha k_{\rm F} L/(\hbar v_{\rm F})$
for the magnetic field and SO interaction, respectively.

\begin{figure}
\begin{center}
\includegraphics[width=88mm]{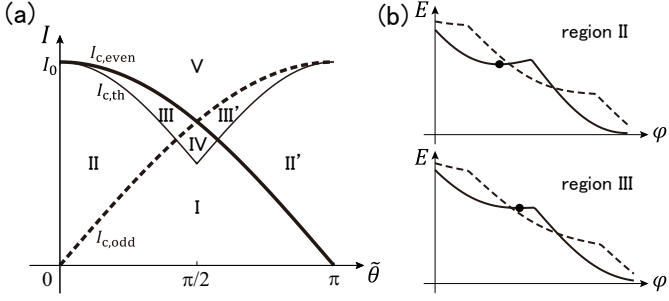}
\end{center}
\caption{
(a) Critical current as a function of the phase $\tilde{\theta}$.
Solid and broken lines indicate the current for the even and
odd parity states, respectively, whereas thin line is one for
the ``thermodynamic'' current.
(b) Schematic view of energy at a finite current bias as
a function of $\varphi$. Solid and broken lines corresponds to
the even and odd parities, respectively. The dot indicates
a stable point for even parity. In the regions II and III,
stable point is absent for odd parity.
}
\label{fig:eoth}
\end{figure}

The ground state energy is given by
$E_{\rm gs} (\varphi) = (1/2) {\sum_n}^\prime E_n (\varphi)$
where the summation is taken over all with negative energy levels.
The ground state is spin singlet in the absence of magnetic field.
In excited states, positive levels are populated~\cite{Chtchelkatchev}.
When the magnetic field is weak,
$E_{\circlearrowright ,\pm}$ and $E_{\circlearrowleft ,\pm}$
are positive and negative at $\varphi =0$, respectively.
We plot schematically the energies of junction with
zero [$E_0 (\varphi) = (E_{\circlearrowleft ,+} + E_{\circlearrowleft ,-})/2$],
one [$E_{1\pm} (\varphi) = E_0 + E_{\circlearrowright ,\pm}$], and
two quasiparticles
[$E_2 (\varphi) = E_0 + (E_{\circlearrowright ,+} + E_{\circlearrowright ,-})$]
in Fig.\ \ref{fig:Model}(c).
The fermion parity of $E_0$ and $E_2$ is even, whereas
$E_{1\pm}$ is odd. At zero temperature, the parity is
conserved and $E_{1\pm}$ states can not relax to
$E_0$ or $E_2$~\cite{Chtchelkatchev}.

The supercurrents via the even and odd states are calculated by
the differential of the energies at $\varphi$,
$I_{\rm p} (\varphi)= (e/\hbar) dE_{\rm p} (\varphi)/d\varphi$
(p $=$ even or odd), where $E_{\rm even}=\min (E_0 , E_2)$ and
$E_{\rm odd}=\min (E_{1+} , E_{1-})$.
Figure \ref{fig:Model}(d) shows schematically the supercurrent
\begin{eqnarray}
I_{\rm even} (\varphi ) &=& \left\{ \begin{array}{c}
I_0 \cos (\tilde{\theta} /2) \sin (\varphi /2)
\hspace{3mm} {\rm for} \hspace{3mm} 0 \leq \varphi < \pi
\\
- I_0 \cos (\tilde{\theta} /2) \sin (\varphi /2)
\hspace{3mm} {\rm for} \hspace{3mm} \pi \leq \varphi < 2\pi
\end{array} \right. , \\
I_{\rm odd} (\varphi ) &=& - I_0 \sin (\tilde{\theta} /2) \cos (\varphi /2)
\end{eqnarray}
with $I_0 \equiv e\Delta_0 /\hbar$. In the presence of
magnetic field, the energy of odd parity, $E_{1+}$, is lower than
the even parity, $E_0$ and $E_2$ at $|\varphi -\pi| < \tilde{\theta}$.
At finite temperature, the parity can change since
the quasiparticle can enter or leave the junction.
If the parity switches immediately, the ``thermodynamic'' current
$I_{\rm th} (\varphi)$ of is determined from $E_{\rm gs}$,
\begin{equation}
I_{\rm th} (\varphi ) = \left\{ \begin{array}{c}
I_0 \cos (\tilde{\theta} /2) \sin (\varphi /2)
\hspace{3mm} {\rm for} \hspace{3mm}
0 \leq \varphi < \pi - \tilde{\theta}
\\
- I_0 \sin (\tilde{\theta} /2) \cos (\varphi /2)
\hspace{3mm} {\rm for} \hspace{3mm}
\pi - \tilde{\theta} \leq \varphi < \pi + \tilde{\theta}
\\
- I_0 \cos (\tilde{\theta} /2) \sin (\varphi /2)
\hspace{3mm} {\rm for} \hspace{3mm}
\pi + \tilde{\theta} \leq \varphi < 2\pi
\end{array} \right..
\end{equation}

\begin{figure}
\begin{center}
\includegraphics[width=80mm]{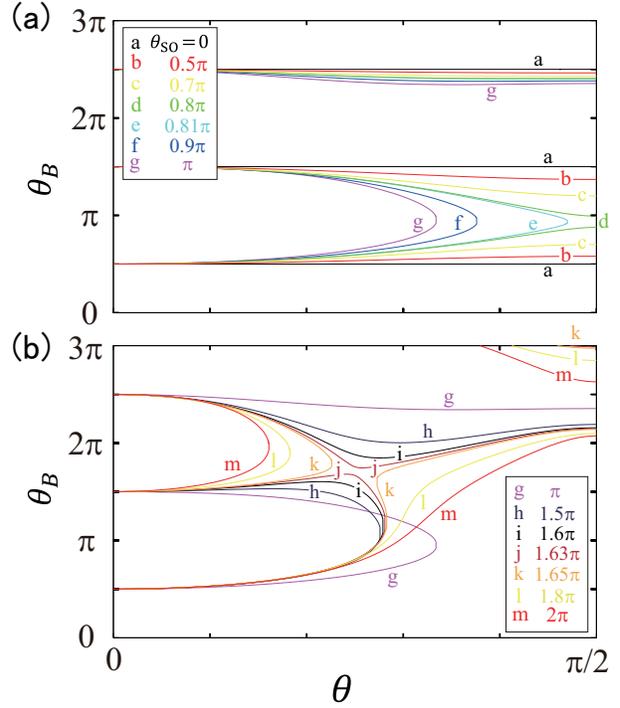}
\end{center}
\caption{
Position of cusps of critical current in the plane of angle $\theta$
and magnetic field $\theta_B = E_{\rm Z} L/(\hbar v_{\rm F})$.
The SO interaction increases from $\theta_{\rm SO} = 0$ to $\pi$ (a)
and $\pi$ to $2\pi$ (b) with
$\theta_{\rm SO} = 2\alpha k_{\rm F} L/(\hbar v_{\rm F})$.
}
\label{fig:0to2}
\end{figure}

The critical currents for even and odd parities and
thermodynamic one are
\begin{eqnarray}
I_{\rm c, even} &=& I_0 \cos (\tilde{\theta} /2)
\\
I_{\rm c, odd} &=& I_0 \sin (\tilde{\theta} /2),
\\
I_{\rm c, th} &=& \left\{ \begin{array}{c}
I_0 \cos^2 (\tilde{\theta} /2)
\hspace{3mm} {\rm for} \hspace{3mm} \tilde{\theta} < \pi /2
\\
I_0 \sin^2 (\tilde{\theta} /2)
\hspace{3mm} {\rm for} \hspace{3mm} \tilde{\theta} \geq \pi /2
\end{array} \right.,
\end{eqnarray}
respectively. At zero temperature, the critical current should be
${\rm max}(I_{\rm c, even},I_{\rm c, odd})$, which is larger than
$I_{\rm c, th}$ [Fig.\ \ref{fig:eoth}(a)]. In Fig.\ \ref{fig:eoth}(a),
we divide the $I$-$\tilde{\theta}$ plane into five regions.
For an applied current $I$ which satisfies $I_{\rm c, even} >I> I_{\rm c, odd}$,
the currents via the even and odd parity state accompany
no and finite voltage, respectively.
The dynamics of phase difference $\varphi$ in the junction at
a finite current is intuitively understood by the so-called tilted
washboard model~\cite{Tinkham}.
The energy for even and odd parities at the current $I$ is given as
$E_{\rm p} (\varphi ,I) = E_{\rm p} (\varphi ) - (\hbar I/2e) \varphi$.
If $E_{\rm p} (\varphi ,I)$ has a stable point, where
$\partial E_{\rm p} / \partial \varphi =0$,
the current accompanies zero voltage [Fig.\ \ref{fig:eoth}(b)].
At finite low temperature, the parity is switched by the thermal
fluctuation. In experiments, the time average of voltage should
indicate a small value, which is determined by the ration between
the dwell time in odd and even states, $\tau_{\rm o}/\tau_{\rm e}$.
In region II in Fig.\ \ref{fig:eoth}(a), $\tau_{\rm o}/\tau_{\rm e} \ll 1$
since the stable point of even parity state is energetically
lower than the energy of odd one at the same phase difference
$\varphi$ [upper panel in Fig.\ \ref{fig:eoth}(b)].
In region III, the stable point of even parity is not energetically
favorable (lower panel) and $\tau_{\rm o}/\tau_{\rm e}$ is larger.
The regions are distinguished by the voltage measurement and
its temperature dependence.

The critical current is determined by the phase $\tilde{\theta}$
and the position of cusp is located at $\tilde{\theta} =\pi/2$ for
both ${\rm max}(I_{\rm c, even},I_{\rm c, odd})$ and $I_{\rm c, th}$.
The $0$-$\pi$ transition also takes place at $\tilde{\theta} =\pi/2$.
Therefore we focus on the position of cusp and does not take
care about the parity effect on the critical current oscillation.
When the external magnetic field is parallel with the effective SO field,
$\cos \tilde{\theta} = \cos \theta_B$ and the cusp positions are 
periodic in magnetic field corresponding to $\theta_B = (2n+1)\pi /2$ with
an integer $n$. The SO interaction modulates the critical current
when $\theta \ne 0$. Figure \ref{fig:0to2} shows the position of
cusps as the magnetic field is rotated. The distance of the first and
the second cusps shortens with increasing of $\theta_{\rm SO}$.
The disappearance of two cusps (and $0$-$\pi$ transition) takes
place at $\theta_{\rm SO} > 0.8\pi$. If the SO interaction is stronger,
the second and the third cusps are closer to each other and vanish.
In that case, the first cusp survives for all $\theta$.

\begin{figure}
\begin{center}
\includegraphics[width=70mm]{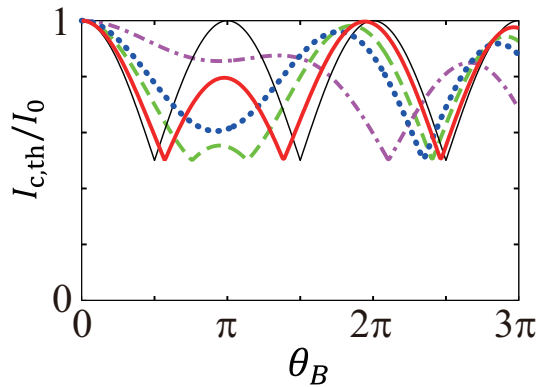}
\end{center}
\caption{
Critical current $I_{\rm c, th}$ as a function of external magnetic
field when $\theta =0.4\pi$. The strength of SO interaction is
$2\alpha k_{\rm F} L/(\hbar v_{\rm F}) = 0.5\pi$ (solid), $0.8\pi$
(broken), $\pi$ (dotted), and $1.5\pi$ (dotted broken lines).
Thin line corresponds to the absence of SO interaction.
$I_0 \equiv e \Delta_0 /\hbar$.
}
\label{fig:Ic}
\end{figure}

We plot the critical current $I_{\rm c, th}$ at $\theta = 0.4\pi$ as
a function of magnetic field in Fig.\ \ref{fig:Ic}. The plots show clearly
the convergence and annihilation of the first and second cusps with
increase of SO interaction. The position of the third cusp also shifts to
weaker magnetic field. The critical current at $\theta_B \approx \pi$
decreases upon increase of $\theta_{\rm SO}$ since the phase
$\tilde{\theta}$ in this case does not reach $\pi$. The disappearance of
the cusps is induced by the strong suppression of $\tilde{\theta}$.
In the absence of SO interaction, $\bm{n}_+ \cdot \bm{n}_- =1$
and Eq.\ (\ref{eq:phase}) becomes $\cos \tilde{\theta} = \cos (\theta_+ + \theta_-)$.
In the presence of SO interaction, the total magnetic fields for
electron and hole in Eq.\ (\ref{eq:totalB}) are not parallel
($\bm{n}_+ \cdot \bm{n}_- <1$), which results in a cancellation of
phase by $\theta_\pm$ in Eq.\ (\ref{eq:phase}). The phase $\tilde{\theta}$
is generally a nonmonotonic function of $\theta_B$ and a local
minimum of critical current without the cusp is found.

This model can explain the experimental results of differential
resistance $dV/dI$ qualitatively~\cite{private}. In the experiment,
the positions of local minimum of critical current for parallel and
perpendicular magnetic field to the nanowire are different.
By tracing the position for all direction of magnetic field,
the orientation of effective field can be evaluated.

We have demonstrated a numerical simulation with
Rashba SO interaction, the result of which agrees with
this simple model~\cite{YEN3}. In this model,
the anomalous Josephson current, $I(\varphi =0) \ne 0$,
is not obtained even in the presence of SO interaction.
The anomalous effect is caused by the spin-dependent
channel mixing due to the SO interaction in previous
studies~\cite{YEN1,YEN2}. If we take into account more than
one conduction channel, the anomalous effect is found.

The anisotropy of g-factor in the nanowire also contributes to
the anisotropy of critical current oscillation. However, the g-factor
anisotropy only shifts the position of cusps and does not induce
the disappearance of $0$-$\pi$ transition. In the experiment for
nanowire quantum dots, the g-factor is the largest for a parallel
direction to the nanowire~\cite{Nadj-Perge2}, which would
shorten the oscillation period although the period is the longest
for the parallel magnetic field in the experiment.
Gharavi {\it et al.} have examined the critical current oscillation in
InAs nanowire, which is attributed not to the Zeeman effect
but to the orbital effect~\cite{Gharavi}.
The orbital effect may contribute to the anisotropy.

\section{CONCLUSIONS}

In conclusions, we have studied the effect of SO interaction on
the critical current in semiconductor nanowire Josephson junction
within a minimal model. The critical current oscillates as a function of
magnetic field. The effective field due to the SO interaction causes
the magnetic anisotropy. We focus on the position of cusps of
critical current that signal the $0$-$\pi$ transition. The oscillation and
its anisotropy are governed by a single phase $\tilde{\theta}$ that
combines the effect of magnetic field and SO interaction. We have also
considered the parity effect on the Josephson current at low temperature.
Although the parity conservation changes the supercurrent and
the character of superconducting transport, it does not affect
the positions of the cusps.

\acknowledgments
We acknowledge fruitful discussions about experiments with
Professor L.\ P.\ Kouwenhoven, A.\ Geresdi, V.\ Mourik, K.\ Zuo of
Delft University of Technology and about theory with
G.\ Campagnano, P.\ Lucignano, Professor A.\ Tagliacozzo of
the University of Naples and D.\ Giuliano of CNR-SPIN.
We acknowledge financial support by the Motizuki Fund of
Yukawa Memorial Foundation.
T.Y. is a JSPS Postdoctoral Fellow for Research Abroad.

\clearpage

\section{\huge{Supplementary note}}

In this supplementary note, we introduce a detail calculation of
scattering-matrix approach in our minimal model.
We also explain the parity switching on the tilted washboard model
and some additional results to support an understanding of physics.

\section{Scattering-matrix in the normal region}

The BdG equation for the envelope function is given by Eq.\ (3) in
the main text:
\begin{equation}
\left( \hspace{-1mm} \begin{array}{cc}
\mp i \hbar v_{\rm F} \partial_x - \bm{h}_\pm \cdot \hat{\bm{\sigma}}
& \Delta (x) \\
\Delta^* (x) &
\pm i \hbar v_{\rm F} \partial_x - \bm{h}_\mp \cdot \hat{\bm{\sigma}} 
\hspace{-1mm} \end{array} \right)
\left( \hspace{-1mm} \begin{array}{c}
\bm{\psi}_{\rm e}^{(\pm)} \\
\bm{\phi}_{\rm h}^{(\pm)}
\end{array} \hspace{-1mm} \right)
= E
\left( \hspace{-1mm} \begin{array}{c}
\bm{\psi}_{\rm e}^{(\pm)} \\
\bm{\phi}_{\rm h}^{(\pm)}
\end{array} \hspace{-1mm} \right)
\end{equation}
with
\begin{equation}
\bm{h}_\pm = \frac{1}{2} E_{\rm Z} \bm{e}_z
\pm \alpha k_{\rm F} \bm{e}_\theta,
\end{equation}
which means a total magnetic field for electron and hole. Let us
concentrate on the right-going electron $\bm{\psi}_{\rm e}^{(+)} (x)$ in
the normal region ($\Delta =0$). The transmission matrix $\hat{t}_{\rm RL}$
connects the wavefunctions at $x =0$ and $L$,
\begin{equation}
\left( \hspace{-1mm} \begin{array}{c}
\psi_{\rm e+}^{(+)} (L) \\
\psi_{\rm e-}^{(+)} (L)
\end{array} \hspace{-1mm} \right)
= \hat{t}_{\rm RL}
\left( \hspace{-1mm} \begin{array}{c}
\psi_{\rm e+}^{(+)} (0) \\
\psi_{\rm e-}^{(+)} (0)
\end{array} \hspace{-1mm} \right).
\label{eq:tRL}
\end{equation}
The wavefunction $\bm{\psi}_{\rm e}^{(+)} (x)$ obeys a following equation,
\begin{equation}
\partial_x \bm{\psi}_{\rm e}^{(+)} =
+i \frac{E +\bm{h}_+ \cdot \hat{\bm{\sigma}}}{\hbar v_{\rm F}}
\bm{\psi}_{\rm e}^{(+)}
\equiv \hat{f}^{(+)} \bm{\psi}_{\rm e}^{(+)}.
\label{eq:EQM}
\end{equation}
$\hat{f}^{(+)}$ is a $2 \times 2$ matrix in the spin space.
We diagonalize this matrix by an unitary matrix,
$\hat{U}^\dagger \hat{f}^{(+)} \hat{U} = {\rm diag} (\lambda_+,\lambda_-)$.
If we define as $\bm{\xi}_{\rm e}^{(+)} = \hat{U}^\dagger \bm{\psi}_{\rm e}^{(+)}$,
Eq.\ (\ref{eq:EQM}) is rewritten as
\begin{equation}
\partial_x
\left( \hspace{-1mm} \begin{array}{c}
\xi_{\rm e+}^{(+)} (x) \\
\xi_{\rm e-}^{(+)} (x)
\end{array} \hspace{-1mm} \right) =
\left( \hspace{-1mm} \begin{array}{cc}
\lambda_+ & \\
 & \lambda_- 
\end{array} \hspace{-1mm} \right)
\left( \hspace{-1mm} \begin{array}{c}
\xi_{\rm e+}^{(+)} (x) \\
\xi_{\rm e-}^{(+)} (x)
\end{array} \hspace{-1mm} \right).
\end{equation}
The solution of this equation,
$\xi_{\rm e \pm}^{(+)} (x) \propto e^{\lambda_\pm x}$, results in
\begin{equation}
\bm{\psi}_{\rm e}^{(+)} (L) = \hat{U}
\left( \hspace{-1mm} \begin{array}{cc}
e^{\lambda_+ L} & \\
 & e^{\lambda_- L}
\end{array} \hspace{-1mm} \right)
\hat{U}^\dagger \bm{\psi}_{\rm e}^{(+)} (0).
\label{eq:psiL0}
\end{equation}
By comparing Eq.\ (\ref{eq:tRL}) and (\ref{eq:psiL0}),
we obtain the transmission matrix $\hat{t}_{\rm RL}$ in
Eq.\ (6) in the main text,
\begin{equation}
\hat{t}_{\rm RL} = \exp
\left( i \frac{L}{\hbar v_{\rm F}} \bm{h}_+ \cdot \hat{\bm{\sigma}} \right),
\end{equation}
where the energy dependent term is neglected.
The transmission matrix for left-going electron,
$\hat{t}_{\rm LR}$, is calculated in the same way.
The scattering-matrix for hole is also obtained directly
from the equation of $\bm{\phi}_{\rm h}^{(\pm)}$,
which satisfies the relation
$\hat{S}_{\rm h} = \hat{g} \hat{S}_{\rm e}^* \hat{g}^\dagger$.

\section{Andreev reflection with spin-orbit interaction}

The Andreev reflection is also expressed in terms of scattering-matrix,
the expression of which is developed by Beenakker~\cite{Beenakker2}.
Beenakker formulates $\hat{r}_{\rm he}$ and $\hat{r}_{\rm eh}$ in
the absence of SO interaction. We consider the Andreev reflection
with SO interaction and show that the SO interaction does not
affect the Andreev reflection coefficient.

We assume SN interface at $x=0$. The superconducting and normal
regions are $x>0$ and $x<0$, respectively. No magnetic field is applied.
For $|E|<\Delta_0$ in the superconducting region, the envelope
function of $\Psi (x) = e^{+ik_{\rm F} x} \tilde{\Psi} (x)$ decays exponentially,
$\tilde{\Psi} (x) = (f_{\rm e},f_{\rm h})^{\rm T} \exp[-\kappa x/\hbar v_{\rm F}]$.
By substituting $\tilde{\Psi} (x)$ into the BdG equation,
\begin{equation}
\left( \hspace{-1mm} \begin{array}{cc}
+i\kappa  -\bm{h}_{\rm SO} \cdot \hat{\bm{\sigma}} -E& \Delta \\
\Delta^* & -i\kappa +\bm{h}_{\rm SO} \cdot \hat{\bm{\sigma}} -E
\hspace{-1mm} \end{array} \right)
\left( \hspace{-1mm} \begin{array}{c}
f_{\rm e} \\
f_{\rm h}
\end{array} \hspace{-1mm} \right) = 0.
\end{equation}
The solution is $\kappa =-\sqrt{\Delta_0^2 - E^2} \pm i |\bm{h}_{\rm SO}|$.
Thus the SO interaction induces the oscillation component in
the evanescent wave. The spin quantization axis is taken in
the direction of $\bm{h}_{\rm SO}$. The Andreev reflection
coefficient converting electron to hole is given as
\begin{equation}
f_{\rm h}/f_{\rm e} = \frac{\Delta^*}{E + i\sqrt{\Delta_0^2 - E^2}}
= \exp \left( -\varphi - \arccos \frac{E}{\Delta_0} \right),
\end{equation}
which is the same result as that without SO interaction.

\section{Parity switching}

At zero temperature, the quasiparticle fermion parity is conserved.
Thus the junction energy or supercurrent for even and odd parity
are taken into account. The junction energies with zero, one, and
two quasiparticles are given in the main text as
\begin{eqnarray}
E_0 (\varphi) &=& -\Delta_0 \cos (\tilde{\theta}/2) \cos (\varphi /2), \\
E_{1\pm} (\varphi) &=& \mp\Delta_0 \sin (\tilde{\theta}/2) \sin (\varphi /2), \\
E_2 (\varphi) &=& \Delta_0 \cos (\tilde{\theta}/2) \cos (\varphi /2),
\end{eqnarray}
respectively. Two quasiparticles can enter into or
leave from the junction immediately if the parity does not change.
The energies for even and odd parity are $E_{\rm even} (\varphi)=\min (E_0 , E_2)$
and $E_{\rm odd} (\varphi)=\min (E_{1+} , E_{1-})$, respectively.
The supercurrents via even and odd parity state in Eqs.\ (13)
and (14) in the main text are calculated by the differential of
the energies at $\varphi$,
$I_{\rm p} (\varphi)= (e/\hbar) dE_{\rm p} (\varphi)/d\varphi$
(p $=$ even or odd). When the temperature is enough high,
the parity changes immediately. In this case, the energy of
junction is given by the ground state energy
$E_{\rm gs} (\varphi) = \min (E_0, E_{1\pm}, E_2)$ and
the current becomes $I_{\rm th} (\varphi)$ in Eq.\ (15) in
the main text.

When a bias voltage is applied to the junction, the phase difference
$\varphi$ would evolve according to $d \varphi /dt = 2eV/\hbar$.
In other word, $\varphi$ would be fixed if the voltage is zero and
only the supercurrent flows. The dynamics of phase difference is
understood intuitively by the so-called tilted washboard model~\cite{Tinkham}.
The energy for even and odd parities at the current $I$ is given as
\begin{equation}
E_{\rm p} (\varphi ,I) = E_{\rm p} (\varphi ) - (\hbar I/2e) \varphi.
\end{equation}
The phase difference $\varphi$ runs down on this potential.
If $E_{\rm p} (\varphi ,I)$ has a stable point, which means
$\partial E_{\rm p} / \partial \varphi =0$, $\varphi$ stops at
the stable point and the current flows accompanying zero voltage.
When an applied current $I$ is small, both even and odd parity
states have the stable point. As the current increases at
$\tilde{\theta} < \pi/2$, the stable point for odd parity vanishes,
whereas the even one has, as shown in Fig.\ 2(b) in the main text.
Then, the current satisfies $I_{\rm c, even} >I> I_{\rm c, odd}$.
Since the fermion parity is kept at zero temperature,
the supercurrent is assured by the even parity state.

At finite temperature, the fermion parity can be switched by
the thermal fluctuation, which changes the number of
quasiparticles in the junction. The probability of parity switching
is estimated as $\sim \exp [-\Delta E (\varphi ,I)/(k_{\rm B}T)]$
with the energy difference $\Delta E$ between even and
odd parity states at phase difference $\varphi$.
If a stationary current within $I_{\rm c, even} >I> I_{\rm c, odd}$
is applied [region II or III in Fig.\ 2(a) in the main text],
a finite voltage, $V_{\rm o}$, due to the odd parity would be
detected by the parity switching. ($V_{\rm e} =0$ in this case.)
The time average of measured value of voltage is
$\bar{V} = V_{\rm o} \tau_{\rm o}/(\tau_{\rm o} + \tau_{\rm e})$
with $\tau_{\rm o (e)}$ being the dwell time in odd (even)
parity state. The measurement value would be determined by
a ration of dwell times $\tau_{\rm o}/\tau_{\rm e}$.
In region II, the current is mainly carried via the even parity
state accompanying no voltage, then the phase difference is
fixed at the stable point. When the parity switches to odd by
the thermal fluctuation, the phase difference goes down on
the slope [Fig.\ 2(b) in the main text]. The energies for
the even and odd parity states crosses with each other.
The parity can change to the state with lower energy.
$\varphi$ stops at the stable point after the parity is back to
even. In region II, the even state energy at the stable point is
lower than the energy of odd one at the same phase difference.
Thus, $\varphi$ goes easily back to the stable point, and
the dwell time in the odd parity state is much shorter than
that in the even one, $\tau_{\rm o} \ll \tau_{\rm e}$.
In region III, on the other hand, the stable point is not favorable
energetically, which results a long dwell time in the odd parity.
If the parity switching takes place immediately at high
temperature, $\varphi$ relaxes to the lower energy before
running on the slope, whereas $\varphi$ can not stop in region III.
Thus, the current accompanies no (finite) voltage in region II (III).

\begin{figure}
\begin{center}
\includegraphics[width=75mm]{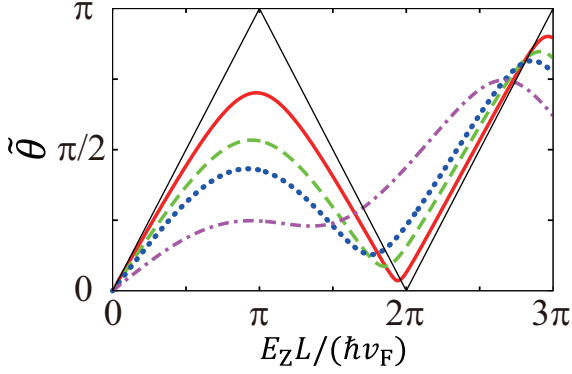}
\end{center}
\caption{
Phase $\tilde{\theta}$ as a function of external magnetic field
when $\theta =0.4\pi$. The strength of SO interaction is
$2\alpha k_{\rm F} L/(\hbar v_{\rm F}) = 0.5\pi$ (solid), $0.8\pi$
(broken), $\pi$ (dotted), and $1.5\pi$ (dotted broken lines).
Thin line corresponds to the absence of SO interaction.
}
\label{fig:SPhase}
\end{figure}
\section{Additional results for magnetic anisotropy}

In the main text, we discuss the magnetic anisotropy of critical
current and $0$-$\pi$ transition. Figure 4 in the main text
demonstrates the critical current oscillation
$I_{{\rm c,th}} [\tilde{\theta} (\theta_B ,\theta_{\rm SO}, \theta)]$
as the magnetic field is applied at $\theta =0.4\pi$ and
the SO interaction is gradually increased. Here we show
additional results to support an understanding of physics.
Figure \ref{fig:SPhase} shows the phase $\tilde{\theta}$ as
a function of magnetic field, where the parameters are
the same as those in Fig.\ 4 in the main text. In the absence of
SO interaction, Eq.\ (12) in the main text becomes
$\cos \tilde{\theta} = \cos \theta_B$. $\tilde{\theta}$
($\in [0.\pi]$) is a sawtooth periodic function of $\theta_B$
and touches $\pi$ at $\theta_B =(2n+1)\pi$ with integer $n$.
When $\tilde{\theta} = \pi/2$, the $0$-$\pi$ transition takes
place. In the presence of SO interaction, $\tilde{\theta}$ is
deviated from the sawtooth behavior and does not reach $\pi$,
which corresponds to the suppression of critical current
$I_{\rm c,th}$. As $\theta_{\rm SO}$ increases, the phase
$\tilde{\theta}$ is suppressed less than $\pi/2$, which
results in the disappearance of $0$-$\pi$ transition around
$\theta_B = \pi$.

\begin{figure}
\begin{center}
\includegraphics[width=75mm]{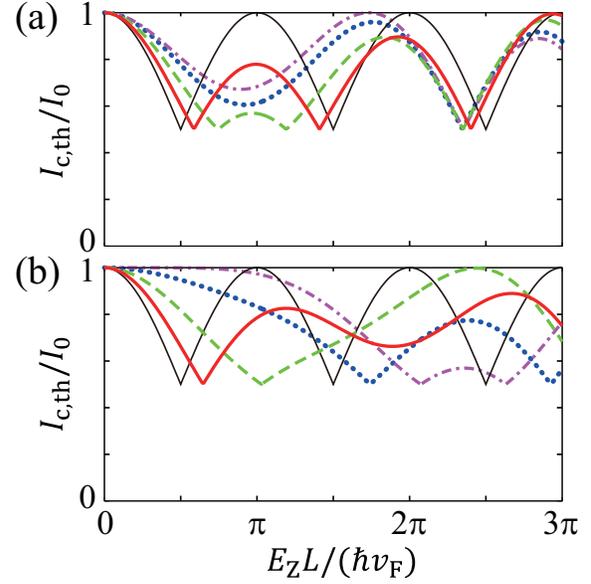}
\end{center}
\caption{
Critical current $I_{\rm c, th}$ as a function of external magnetic
field when the magnetic field is rotated. The strength of SO interaction
is $2\alpha k_{\rm F} L/(\hbar v_{\rm F}) = \pi$ (a) and $2\pi$ (b).
Solid, broken, dotted, and dotted broken lines correspond to
$\theta =0.2\pi$, $0.3\pi$, $0.4\pi$, and $0.5\pi$, respectively.
Thin line is $\theta =0$, where the SO interaction does not affect
the critical current. $I_0 \equiv e \Delta_0 /\hbar$.
}
\label{fig:SIc}
\end{figure}

We plot the critical current oscillation additionally when the SO
interaction is fixed and the angle $\theta$ between magnetic field
and effective SO field is tuned in Fig.\ \ref{fig:SIc}. The critical
current around $\theta_B = \pi$ decreases with an increase of
angle $\theta$. The position of cusps are also modified by
the direction of magnetic field. When $\theta_{\rm SO} = \pi$,
the first and second cusps are closer to each other with
an increase of $\theta$, and vanishes. This behavior is similar with
that in Fig.\ 4 in the main text. When $\theta_{\rm SO} = 2\pi$,
the first cusp shifts to large $\theta_B$ by the angle $\theta$.
These results indicates the magnetic anisotropy of critical current
clearly. The periodicity of critical current gives an information of
direction of effective SO field.

\begin{thebibliography}{0}
\bibitem{com}
We use a mathematical formula for the Pauli matrix,
$e^{i a \bm{n}_a \cdot \hat{\bm{\sigma}} }
e^{i b \bm{n}_b \cdot \hat{\bm{\sigma}} }
= e^{i c \bm{n}_c \cdot \hat{\bm{\sigma}} }$, where
$\cos c = \cos a \cos b - (\bm{n}_a \cdot \bm{n}_b) \sin a \sin b$
and
$\bm{n}_c \sin c = \{\bm{n}_a \sin a \cos b + \bm{n}_b \cos a \sin b
- (\bm{n}_a \times \bm{n}_b) \sin a \sin b \}$.
\end{thebibliography}
\end{document}